\documentclass[%
 reprint,
 amsmath,amssymb,
 aps,nofootinbib
]{revtex4-1}

\usepackage{graphicx}
\usepackage{dcolumn}
\usepackage{bm}
\usepackage{hyperref}
\usepackage{graphicx,epstopdf}
\usepackage{bm}
\usepackage{ wasysym }
\usepackage{placeins}
\DeclareGraphicsExtensions{.ps}
\newcommand{\pt}{\ensuremath{p_{\mathrm{T}}}}

\usepackage{xcolor}
\def\Nfour	{\mathcal{N}\,{=}\,4}
\def\Nc		{N_{\rm c}}

\def\half	{\tfrac {1}{2}}
\def\eq		{\,{=}\,}
\begin{document}

\title{\boldmath Phenomenological implications of asymmetric $AdS_5$ shockwave collision studies for  heavy ion physics}
\author{Berndt M\"uller}
 \email{mueller@phy.duke.edu}
 \affiliation{Department of Physics, Duke University, Durham, NC 27708-0305, USA.}
 
\author{Andreas Rabenstein}%
 \email{andreas.rabenstein@physik.uni-regensburg.de}

\author{Andreas Sch\"afer}%
 \email{andreas.schaefer@physik.uni-regensburg.de}
 \altaffiliation{Physics Department, Brookhaven National Laboratory, Upton, NY 11973, USA}

\author{Sebastian Waeber}%
 \email{sebastian.waeber@physik.uni-regensburg.de}
\affiliation{%
Institute for Theoretical Physics, University of Regensburg,
      D-93040 Regensburg, Germany
}%
\author{Laurence G. Yaffe}%
 \email{yaffe@phys.washington.edu}
\affiliation{%
 Department of Physics, University of Washington, Seattle WA 98195-1560, USA
}%

\date{\today}

\begin{abstract}
This paper discusses possible phenomenological implications
for p+A and A+A collisions of the results of recent numerical
AdS/CFT calculations examining
asymmetric collisions of planar shocks.
In view of the extreme Lorentz contraction,
we model highly relativistic heavy ion collisions (HICs)
as a superposition of collisions between many near-independent
transverse ``pixels'' with differing incident longitudinal momenta.
It was found that also for asymmetric collisions the
hydrodynamization time
is in good approximation a proper time,
just like for symmetric collisions,
depending on the geometric mean of the longitudinally integrated
energy densities of the incident projectiles.
For realistic collisions with fluctuations in the initial energy densities,
these results imply a substantial increase in the hydrodynamization time for highly 
asymmetric pixels. However, even in this case the local hydrodynamization time still 
is significantly smaller than  perturbative results for the thermalization time.
\end{abstract}

\pacs{Valid PACS appear here}
\maketitle


\section{Introduction}

High energy p+p, p+A and A+A collisions at RHIC and LHC address
many interesting questions.
Some fundamental ones concern the relationship between quantum
field theory and hydrodynamics, thermalization,
and between quantum entanglement, decoherence,
and thermodynamic entropy production.
According to our present understanding, within
a short time of order $1-2$ fm/$c$  or less, collision systems reach a
state which can be approximated by a  thermal medium characterized
by local thermodynamic properties.
The microscopic processes leading to this state
are still somewhat controversial, the main problem being that
no controlled calculational technique in QCD is
applicable to the {strongly time dependent rapidly evolving, far off
equilibrium, and strongly coupled medium produced by the collisions.

Although there exist conflicting observations, there
seems to be wide agreement that two complementary approaches
have been quite successful in modeling and analyzing the pre-hydro
($< 0.1-0.2$ fm/$c$) and later ($> 1-2$ fm/$c$) phases of such collisions,
namely AdS/CFT (or gauge/gravity) duality and relativistic viscous
hydrodynamics.
As was shown in Ref.~\cite{vanderSchee:2013pia} for smooth shocks,
i.e.\ neglecting initial state fluctuations,
the interpolation between these two regimes is quite smooth.
This observation suggests that potential
theoretical concerns regarding the use of
AdS/CFT duality to model the strongly coupled dynamics
of QCD plasma prior to the onset of a hydrodynamic regime
are not so severe as to foreclose the phenomenological utility
of this approach.

AdS/CFT duality, in its simplest form, allows one to  solve
the dynamics of maximal supersymmetric $SU(N)$ Yang-Mills theory
($\Nfour$ SYM) in the limit of large $N$ and large 't Hooft coupling.
 Such a system, of course, is not QCD.
Therefore, much energy has been invested in recent years
characterizing the differences in the dynamics
between a real quark-gluon plasma (at accessible temperatures)
and the non-Abelian plasma of $\Nfour$ SYM. 
In Ref.~\cite{Waeber:2018bea} we extended earlier literature
on finite 't Hooft coupling corrections
\cite{Buchel:2008ac,Buchel:2008sh,Stricker:2013lma}
by evaluating corrections to the lowest
electromagnetic quasinormal mode (QNM) frequencies.
The inverse of the
imaginary part of the lowest QNM frequency gives a characteristic
thermalization time and is, therefore, especially relevant for the
dynamics of HICs.
In a different work \cite{Endrodi:2018ikq},
we compared the response of QCD and $\Nfour$ SYM plasmas to
a background magnetic field, and with an appropriately
calibrated comparison found remarkably little difference between
the behavior of QCD and that of conformal $\Nfour$ SYM over
a wide range of temperature and magnetic field.
Lattice gauge theory studies \cite{Panero:2009tv,Bali:2013kia}
have shown that meson masses and some meson
coupling constants scale trivially with the number of colors
all the way from $N=3$ to $N=\infty$.

 In these studies,
modeling QCD plasma at experimentally relevant temperatures
using large-$N$, strongly coupled $\Nfour$ SYM theory
works much better than might have been expected {\em a priori}.
Based on these results, we surmise that strongly coupled
$\Nfour$ SYM, to which AdS/CFT duality is most easily applied,
describes the early phase of HICs not only
qualitatively but also semi-quantitatively at a useful level of accuracy.%
\footnote
    {%
    To be clear, this assumption applies to observables sensitive to
    typical thermal momentum scales in the plasma and not, for example,
    to measurements of high transverse momentum particle or jet production.
    }
This evidence-based hypothesis forms the basis for the considerations
in our present work.

It is remarkable how well viscous hydrodynamics describes HICs.
Various hydrodynamics codes achieve a close to perfect agreement with an
enormous amount of experimental data in spite of uncomfortably large
spatial gradient terms. Considerations that help explain this
unexpected success include the distinction between
``hydrodynamization'' and genuine ``thermalization,''
and the fact that hydrodynamics has attractor properties which set in
long before true local equilibration is reached
 \cite{Berges:2013fga,Romatschke:2017vte,Romatschke:2017acs}.
A rather disquieting
consequence of this line of argument is that while overwhelming
experimental evidence supports hydrodynamic behavior and suggests
early hydrodynamization, there is little experimental evidence for early
genuine thermalization. The latter is, however, required to fulfill the
core  premise of high energy heavy ion physics, namely that 
nuclear collisions allow us to investigate the equilibrated 
quark-gluon plasma that filled the early universe.
We try in this contribution to add one piece of information to
this complicated problem.

\begin{figure}
\includegraphics[width=8.6cm]{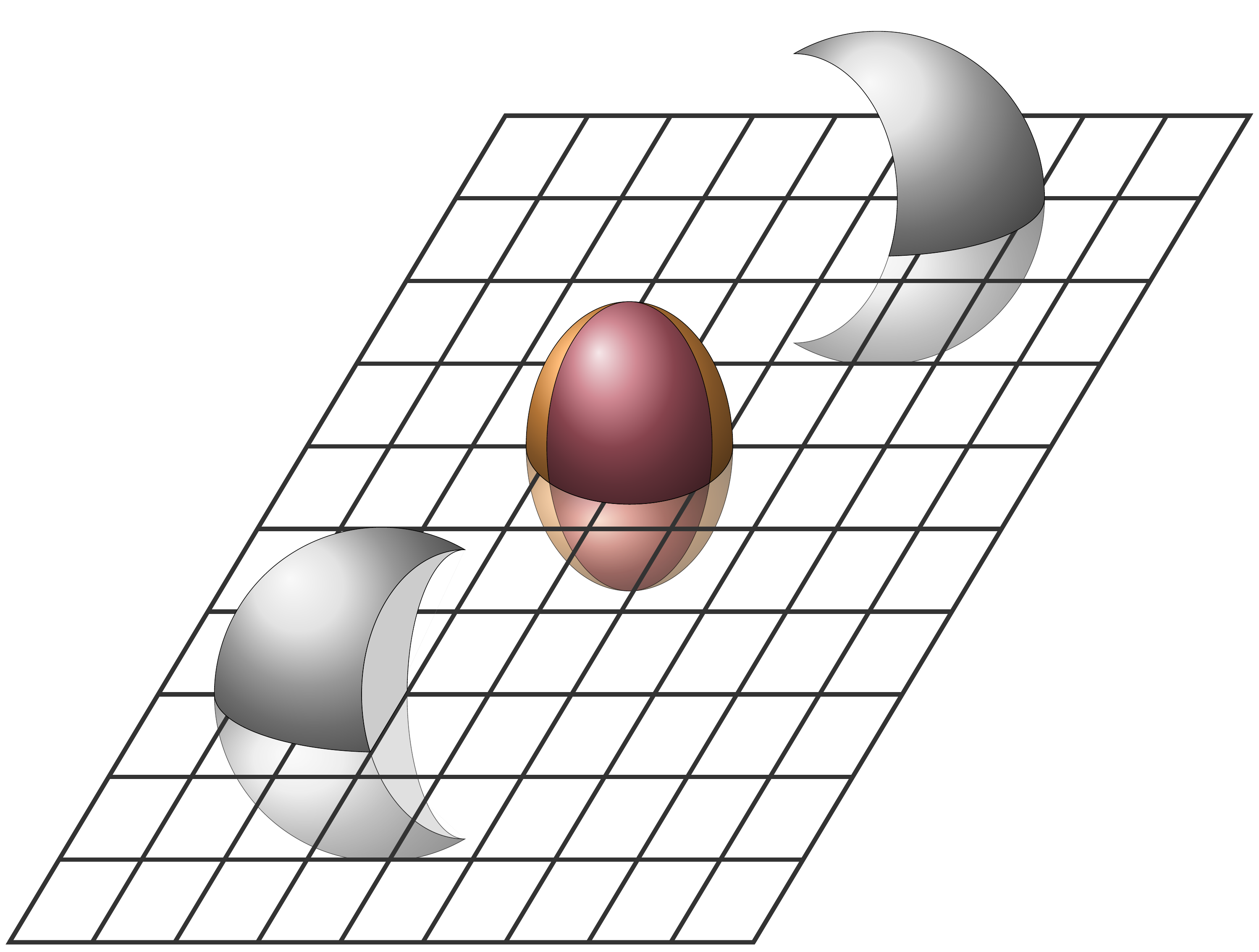}
\caption
    {%
    Sketch of a peripheral HIC.}
\label{fig:collision}
\end{figure}

High energy heavy ion experiments have generated many surprising
experimental observations which call for microscopic explanations.
One is the degree of similarity between high multiplicity p+p collisions
and A+A collisions.
Another surprise is the extent to which
the usual cartoons illustrating HICs,
such as Fig.~\ref{fig:collision}, involving smooth energy densities
for the colliding nuclei are quite misleading.
Instead, modern hydrodynamics codes typically start from initial conditions
with extremely large fluctuations of energy and entropy density.
(See figure~1 in Ref.~\cite{Bernhard:2016tnd} for a typical example.)
These initial conditions are required to
explain the large observed odd azimuthal flow moments,
see e.g., Ref.~\cite{Acharya:2018zuq}.
If odd moments solely arose from statistical
fluctuations of the hydrodynamic fluid itself,
then for symmetric collisions such as Pb+Pb,
the odd flow coefficients $v_3$, $v_5$, etc., should be very
significantly suppressed compared to the even coefficients
$v_2$, $v_4$, etc., which is not the case.
Here, as usual, the flow moments $\{ v_{\rm n} \}$ are defined by
the azimuthal dependence of the produced particle distribution,
\begin{equation}
\label{eq:flowdud}
    E \, \frac{\mathrm{d}^{3}N}{\mathrm{d}p^3}
    =
    \frac{1}{2\pi} \,
    \frac{\mathrm{d}^{2}N}{p_{\rm T}\, \mathrm{d} \pt{}\, \mathrm{d} y}
    \Bigl(1+2\sum_{\rm n=1}^{\infty}v_{\rm n} \cos[{\rm n}(\varphi - \Psi_{\rm n})]\Bigr),
\end{equation}
with $E$ the energy, $p$ momentum, \pt{} transverse
momentum, $\varphi$ the azimuthal angle, $y$ the pseudorapidity
of the particle, and $\Psi_{\rm n}$ the $n$-th harmonic symmetry
plane angle.

Hence typical AdS shock wave calculations involving smooth
initial energy densities are too idealized.
Even symmetric Pb+Pb collisions should be characterized by
initial energy densities which are asymmetric owing to the
presence of independent and substantial transverse variations.
This was one of the motivations for our study \cite{Waeber:2019nqd}
of idealized but highly asymmetric planar shock collisions.

A further phenomenologically relevant aspect is that
AdS/CFT models of collisions, in the leading infinite coupling limit,
tend to predict surprisingly short hydrodynamization times and equilibration  times\footnote{The here cited model treated in \cite{Balasubramanian:2010ce} doesn't sharply differentiate between hydrodynamization time and equilibration time. Here and henceforth we also use the term "equilibration" synonymous with "thermalization".},
0.3 fm/$c$ or less \cite{Balasubramanian:2010ce}.
 While there is no consensus among perturbative, i.e. weak coupling, estimates of the thermalization time scale $\tau_{\rm therm}$, all estimates suggest $\tau_{\rm therm}$ it to be substantially longer. 
For example, an early result \cite{Baier:2000sb} gave
$\tau_{\rm therm}>\alpha^{-13/5}Q_s^{-1}>4$~fm/$c$, 
and a more recent result based on a different analysis \cite{Berges:2014yta} 
is $\tau_{\rm therm}\sim 2$~fm/$c$.
In \cite{Romatschke:2016hle} it is argued from the perspective of hydrodynamics 
that thermalization might possibly never be reached in a realistic heavy-ion collision.

In our earlier work \cite{Waeber:2018bea} using a specific resummation {\em ansatz},
we found that for realistic finite values of the 't Hooft coupling 
of QCD, the imaginary part of the lowest QNM is roughly halved, 
corresponding to a doubling of the predicted equilibration time.
In the present contribution we will argue that for fluctuations
large enough to generate the observed $v_3$ values,
the AdS prediction for the hydrodynamization time of individual transverse ``pixels'' are
significantly lengthened
but stay much smaller than $\tau_{\rm therm}$ found in perturbative calculations. However, we will
also argue that complete thermalization could take much longer as it
requires equilibration between these ``pixels''.

Over the years many different hydrodynamics codes have been developed, 
improved and fine-tuned to describe the experimental data. Their relative
advantages and disadvantages are the topic of specialized workshops.
We do not want to enter this discussion here. Rather, we will focus on
just one relatively recent study \cite{Bernhard:2016tnd} which is especially 
systematic with respect to the properties we are interested in. 
We leave it to the authors of other studies to decide whether our conclusions 
are also relevant for their work.

In the following section we briefly review those results of
Ref.~\cite{Bernhard:2016tnd} which are important for us
and then discuss how these compare with the results of our
AdS/CFT study \cite{Waeber:2019nqd}.
The present contribution was separated from
Ref.~\cite{Waeber:2019nqd} because, unlike the
well defined results of Ref.~\cite{Waeber:2019nqd} which
should stand the test of time,
the following discussion of phenomenological consequences
depends crucially on the comparison of results from hydrodynamics
codes to experimental data and is subject to
far more uncertain interpretation.
In particular, it is not yet feasible to perform numerical
gravity calculations with initial conditions which fully
mimic the strong transverse fluctuations of energy and entropy densities
that appear to be present in real collisions.

In Section 3 we discuss implications for peripheral collisions
and for p+A collisions.  Section 4 is devoted to another aspect,
namely the time dependence of the apparent horizon in asymmetric
collisions and a comparison to what is known about the time dependence
of entropy for classical and quantum gauge theories.
A final section holds a few concluding remarks.

\section{The role of fluctuations in heavy ion collisions}

The arguments in favor of strong fluctuations in the initial state
of heavy ion collisions (HICs) are manifold, both theoretical and experimental.
In, e.g., Ref.~\cite{Muller:2011bb} (see also~\cite{Lappi:2017skr}) the initial fluctuations
in transverse energy density were calculated in the Color Glass
Condensate (CGC) model.
It was argued that for typical pixels these can be larger than 50\%.
On the experimental side we have
already noted the surprisingly large values $v_3$ observed,
e.g., in Ref.~\cite{Acharya:2018zuq}.

Different models vary in their assumptions, including
those which concern fluctuations.
We will follow Ref.~\cite{Bernhard:2016tnd}
whose Fig.~1 shows several typical examples of initial state fluctuations.
The basic assumption of that paper, which we also adopt,
is that one may think of the initial state of a HIC as arising from a
sum over many isolated collisions of often vastly asymmetric ``pixels''.
Because these asymmetries are so large, holographic calculations for
smooth symmetric shock wave collisions are insufficient.

 Extending the AdS treatment to include realistic fluctuations is somewhat subtle because
it relates to basic questions of what is exactly meant by ``decoherence'' and ``thermalization.''
While the fundamental T-invariance of QCD seems to imply the absence of any decoherence, this
is no longer true if specific probes of only limited spatial extent are considered. All standard
observables for high energy heavy-ion collisions do exactly that, probing only transverse scales
which are much smaller than the nuclear radii, be it $1/Q_s$ or, e.g., individual hadron radii.
Therefore, real life heavy ion experiments always  imply coarse graining,
see again  Fig.~1 of
\cite{Bernhard:2016tnd}, which  circumvents this T-invariance argument.
Basically, all experimental observables are insensitive to quantum correlations beyond
the scales mentioned above.

The description of detailed properties of collisions of highly
non-uniform nuclei by viscous relativistic hydrodynamics has been
the topic of many careful and interesting investigations, far
too many to review in this short note. Let us only mention
Ref.~\cite{Welsh:2016siu}, where it was highlighted that the inclusion
of realistic fluctuations is even more important if one studies
collision systems like p+A. The fact that we limit our discussion
here to Ref.~\cite{Bernhard:2016tnd} should not be interpreted as any
form of judgment on the relative value of the various models but
just as reflection of our inability to do justice to all of them.

One of the standard procedures, also adopted here, is to describe
all collisions by means of a ``nuclear thickness function'' $T$
which is usually assumed to be a superposition of Gaussians of a
certain width $w$ in the transverse plane.
In Ref.~\cite{Bernhard:2016tnd}
the randomly generated participant thickness function $\widetilde T(x,y)$ is
constructed as follows:
\begin{subequations}
\begin{eqnarray}
\widetilde T(x,y) &=& \sum_{i=1}^{N_{\rm part}}~~\gamma_i~T_p(x{-}x_i,y{-}y_i)
\\
T_p(x,y)&=& \frac{1}{2\pi w^2} \, \exp\left( -\frac{x^2+y^2}{2w^2} \right)
\end{eqnarray}
\label{eq:Ttilde}%
\end{subequations}
where the coefficients $\gamma_i$ are chosen according to the Gamma
probability distribution,
\begin{equation}
    P(\gamma)
    = \frac{k^k}{\Gamma(k)} \, \gamma^{k-1}e^{-k\gamma} \,,
\label{eq:2.2}
\end{equation}
with the parameter $k=1.4$ and the mean value of $\gamma$ set to unity.
The coefficients $\{\gamma_i\}$ simulate
the statistical fluctuations of the initial state,
while $\{ x_i, y_i \}$ are the random participant locations
in the transverse plane. (See Ref.~\cite{Bernhard:2016tnd} for details.)

When we refer to ``pixels'' we mean independent transverse areas 
with a radius of order $1/Q_s$, where $Q_s$ is the saturation scale
of order 1--2 GeV in the initial state and not those areas with
which hydrodynamics is initialized (with mean radius $w$), which
we call ``patches`` for distinction.
The typical radius of a ``pixel''
is 0.1--0.2 fm, while that of a hydrodynamical ``patch'' is 0.4--1.2 fm,
see table 1 in Ref.~\cite{Moreland:2018gsh}.
However, in line
with our comment above regarding the attractor properties of hydrodynamics,
the initialization time for hydrodynamics can be chosen more or
less at will in the range $0.1-1.5$ fm/$c$
(see again Table 1 in Ref.~\cite{Moreland:2018gsh}).
In Ref.~\cite{Bernhard:2016tnd} the hydrodynamics code
was actually initialized at $t \eq 0$~fm/$c$ which resulted in the fit
value $w\eq 0.5$~fm. This value for the patch size $w$ probably has little
physical meaning, as hydrodynamics is definitely not applicable at
$t \eq 0$~fm/$c$, but this is irrelevant for our discussion since
the distribution $P(\gamma)$ is independent
of $w$ and so are the resulting fluctuations.
We will argue below that the relevant time scales for hydrodynamization and thermalization 
depend only on the distribution function $P(\gamma)$.

The physical idea behind the thickness function is that due to
length contraction and time dilation partons in the colliding nuclei are coherent in
the longitudinal direction but incoherent in transverse distance beyond
a characteristic length scale which is typically chosen as the inverse
saturation scale $1/Q_s\sim 0.2$~fm.
What is phenomenologically
important is the assumption made about how the local initial entropy 
density $s$ (or related energy density) depends on the thickness functions of
two colliding transverse pixels.
Very little is known about this and models differ widely.
The authors of Ref.~\cite{Bernhard:2016tnd}, therefore, use the
flexible parameterisation
\begin{equation}
s(x,y) ~\sim~ \left( \frac{\widetilde T^p_A +\widetilde T^p_B}{2}\right)^{1/p}  
\end{equation}
which covers a wide range of possibilities.
\begin{equation}
    s~\sim~ \left\{
    \begin{array}{lll}
	\max(\widetilde T_A ,\widetilde T_B) \,, & p\rightarrow +\infty;&\\[2pt]
	(\widetilde T_A +\widetilde T_B)/2 \,, & p=+1;& ({\rm arithmetic})\\[2pt]
	(\widetilde T_A \, \widetilde T_B)^{1/2} \,, & p=0;& ({\rm geometric})\\[2pt]
	2(\widetilde T^{-1}_A +\widetilde T^{-1}_B)^{-1} \,, 
	~~~~~& p=-1;~~~~~& ({\rm harmonic})\\[2pt]
	\min(\widetilde T_A ,\widetilde T_B) \,, & p\rightarrow -\infty,&
    \end{array}
\right.
\end{equation}
and simply vary the value of $p$ to find the one for which they obtain the best
overall agreement with the data.
This phenomenological analysis clearly favors the geometric mean ($p=0$),
see Fig.~9 of Ref.~\cite{Bernhard:2016tnd}, so that
\begin{equation}
    s(x,y) \sim \left[ \widetilde T_A(x,y) \,\widetilde T_B(x,y)\right]^{1/2}
    \,.
\label{eq:s(x,y)}
\end{equation}

This dependence of the initial entropy density on the geometric mean
of the thickness functions $\widetilde T_A, \widetilde T_B$
is especially noteworthy in light of analogous results from
studies of holographic collisions \cite{Waeber:2019nqd}.
This study found that in asymmetric collisions
the rapidity dependence of the proper energy density
of the produced plasma, near the onset of the hydrodynamic regime,
is well described by the geometric mean of the produced proper energy densities
in the corresponding symmetric collisions.
Moreover, the hydrodynamization time hypersurface
was found to be an almost perfect proper time hypersurface
(i.e., the boundary of the hydrodynamic regime is essentially a hyperbola)
whose value depends exclusively on the energy scale set by the
geometric mean of the longitudinally integrated energy densities
of the colliding pixels \cite{Waeber:2019nqd ,Chesler:2015fpa},%
\footnote
    {%
    Explicitly, $\int dz \> T^{00}_A(z) \equiv \mu_A^3 \, \Nc^2/(2\pi^2)$,
    etc.
    }
\begin{equation}
    \tau_{\rm hydro} \approx 2/\sqrt{\mu_A \, \mu_B} \,.
\label{eq:thydro}
\end{equation}

\begin{figure}
\includegraphics[width=8.6cm]{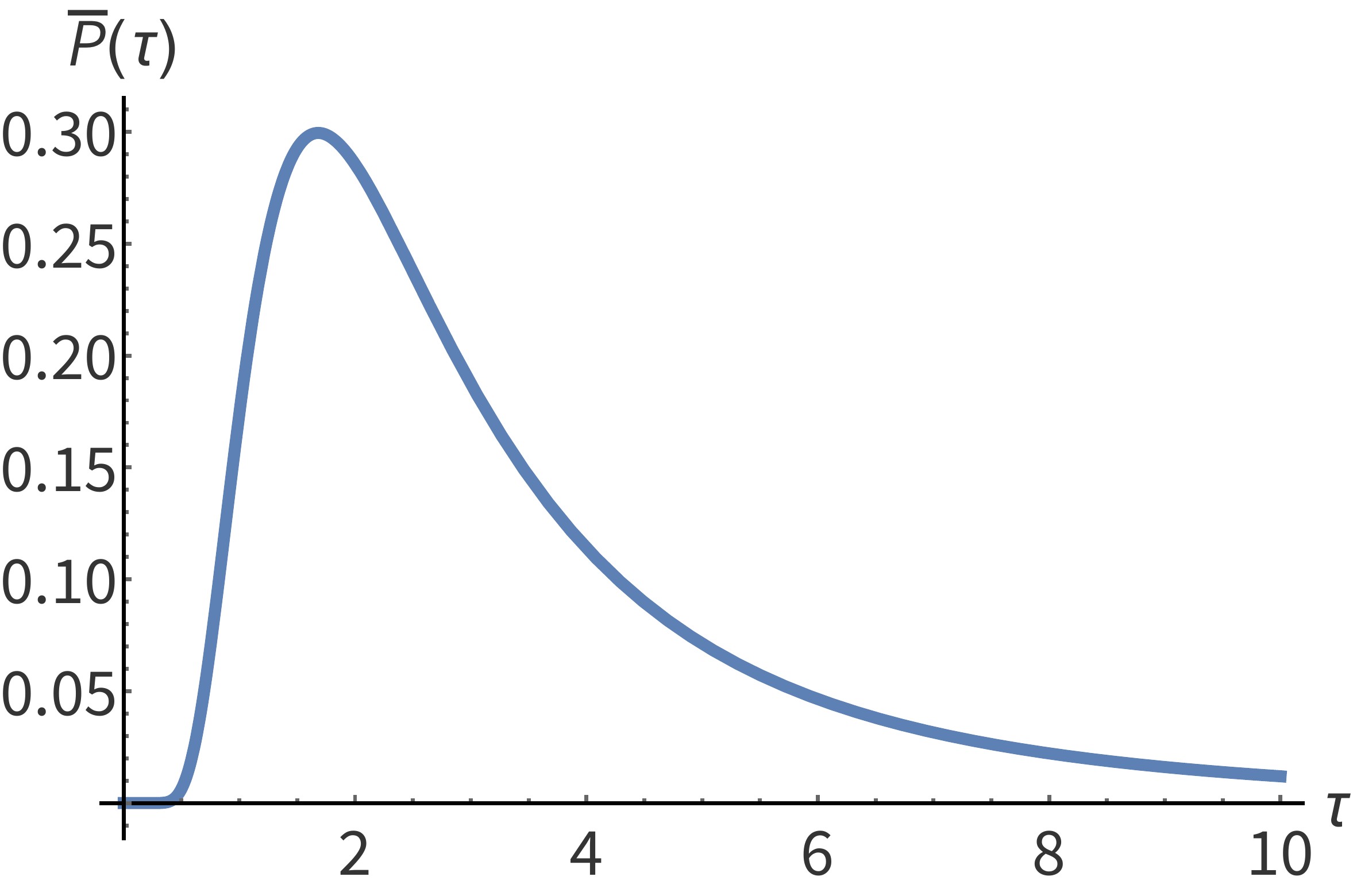}
\caption
    {%
    The probability distribution $\overline P(\tau)$
    for $k=1.4$, in units where $\mu = 1$.
    }
\label{fig:Pbar}
\end{figure}

To connect this holographic result for hydrodynamization time to the
model (\ref{eq:s(x,y)}) of fluctuating initial conditions,
we regard the energy scale $\mu_A$ of a given pixel of projectile $A$
as equal to the average scale $\mu$ times the fluctuating amplitude
$\gamma_i$ of some participant lying within this pixel,
and likewise for projectile $B$.
One may then compute the resulting probability distribution 
$\overline P(\tau_{\rm hydro})$
of the hydrodynamization time $\tau_{\rm hydro}$.
The result is 
\begin{eqnarray}
    &\overline P(\tau)
    \equiv
    \int_0^\infty \int_0^\infty d\gamma_A \> d\gamma_B \>
    P(\gamma_A) \, P(\gamma_B) \,\times \nonumber
\\ &\delta(\tau - \tfrac 2{\mu \sqrt{\gamma_A \gamma_B}})=
    \frac 4{\tau}
    \left(\frac {2k}{\mu\tau} \right)^{2k}
    J_0\Big(\frac{4k}{\mu\tau}\Big) \bigm/
    \Gamma(k)^2 \,.
\end{eqnarray}
This distribution in plotted in Fig.~\ref{fig:Pbar}.
It is peaked at $\tau = 1.7/\mu$, a little below the value
of $2/\mu$ which, from Eq.~(\ref{eq:thydro}), would be the
holographic prediction in the absence of fluctuations.
Glancing at Fig.~\ref{fig:Pbar} and focusing on the peak
in the distribution, one might think that the
main effect of including initial state fluctuations is merely
to induce relatively modest fluctuations in the hydrodynamization
time so that it becomes, crudely,
$\tau_{\rm hydro} \approx (2\pm 1)/\mu$.
This conclusion, however, is wrong due to the slow
power-law decrease of the distribution $\overline P$ with increasing
time $\tau$, which reflects the non-analytic behavior
of the adopted form (\ref{eq:2.2}) of 
$P(\gamma)$ as $\gamma \to 0$.

The median of the distribution for $\tau_{\rm hydro}$
is $2.76/\mu$ (for $k = 1.4$),
substantially larger than the peak value.
The mean hydrodynamization time is given by
\begin{subequations}
\begin{eqnarray}
    \bar \tau_{\rm hydro}
    \equiv
    \int_0^\infty d\tau \> \overline P(\tau) \, \tau
    &=&
    \frac{2k}{\mu} \, \frac{\Gamma(k{-}\half)^2}{\Gamma(k)^2}
\\ &=&
    4.06 / \mu \,,
\end{eqnarray}
\end{subequations}
with the numeric value specific to $k = 1.4$,
while the rms deviation is
\begin{subequations}
\begin{eqnarray}
    \Delta \bar \tau_{\rm hydro}
    &\equiv &
    \left[
    \int_0^\infty d\tau \> \overline P(\tau) \,
    (\tau - \bar\tau_{\rm hydro})^2
    \right]^{1/2}
\\  &=&
    \frac {2k}{(k{-}1) \,\mu}=
    5.70 / \mu \,,
\end{eqnarray}
\end{subequations}
considerably larger than the mean value.
There is a 70\% probability that the hydrodynamization time
is larger than the non-fluctuating estimate of $2/\mu$,
a 45\% probability that $\tau_{\rm hydro}$ is larger than $3/\mu$,
and a 30\% probability that it is larger than $4/\mu$.


This simple model predicts that fluctuations in the transverse plane
lead to large non-Gaussian fluctuations in the hydrodynamization time.
For an individual collision of two pixels it doubles, on
average, the hydrodynamization time,
thereby converting the typical time scale of $0.2-0.3$ fm/$c$ predicted
by AdS/CFT modeling with smooth initial conditions
\cite{Balasubramanian:2010ce} 
to the range $0.4-0.6$ fm/$c$.
Moreover, it has consequences for
thermalization of the complete nuclear system. Complete 
equilibration would require equilibration between different
pixels.
However, with significant variations in energy density and hence
effective local temperature across the transverse extent of the fireball,
such variations can only relax via hydrodynamic processes whose
time scale grows linearly with the length scale of variations and
exceeds the relevant initial hydrodynamization time.
The fact that fluctuation are definitely not smoothed out at the initial
hydrodynamization time explains why large fluctuations can still persist
around 1 fm/$c$ and  generate a sizable triangular flow component $v_3$.
In other words, hadron phenomenology as encoded in hydrodynamics codes like
\cite{Bernhard:2016tnd} appears to be compatible with AdS/CFT modeling
only because of large transverse fluctuations.

\section{Peripheral collisions}

\begin{figure}
\includegraphics[width=8.6cm]{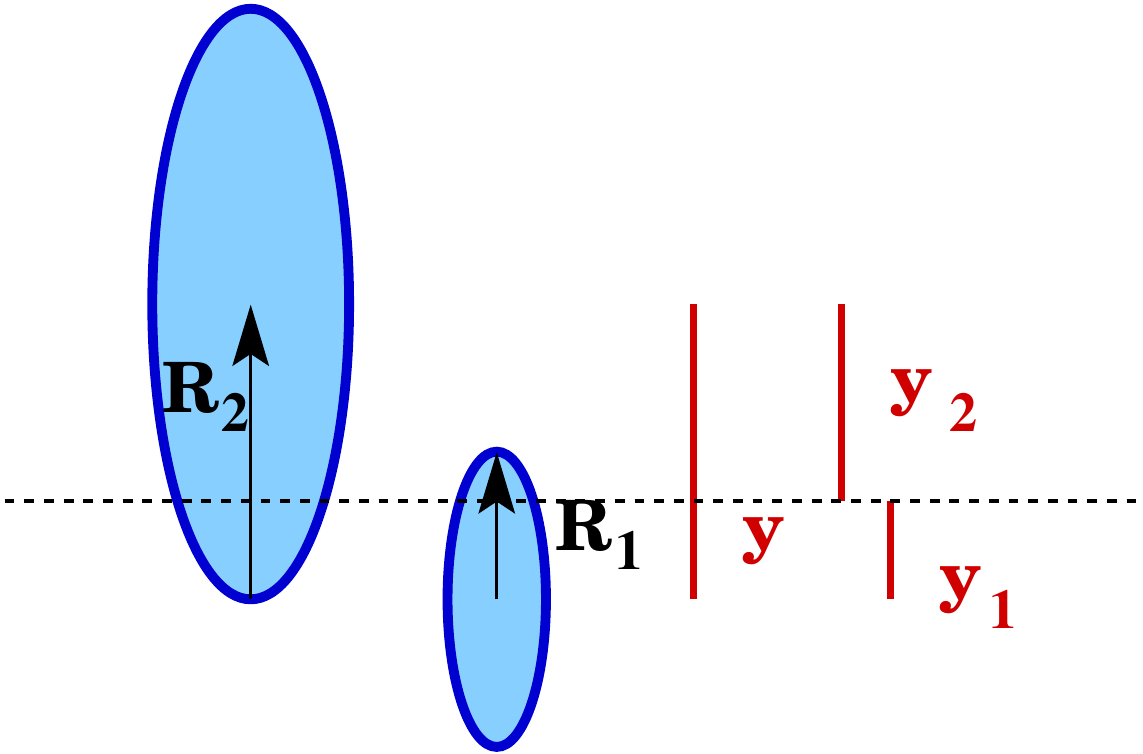}
\caption
    {Sketch of a peripheral heavy ion collision.}
\label{fig:collision2}
\end{figure}

One original motivation for our investigation was the observation that,
if hydrodynamization would occur significantly more slowly in the
asymmetric fringe regions (the orange areas in Fig.~\ref{fig:collision})
in peripheral collisions, then this would cause the hydrodynamized overlap region
(the inner red region in Fig.~\ref{fig:collision}) to be slimmer and thus increase $v_2$.
It turns out, however, that this effect is rather small and can be
essentially neglected in comparison with the large fluctuation effects
just discussed.
To illustrate this, we consider in this section a very crude model
that approximates the energy and entropy densities of each
colliding nucleus as homogeneous within its Lorentz contracted
spherical volume, such that the
asymmetry for a given pixel is exclusively determined by geometry.
The nuclei move in the $\pm z$ direction and the reaction plane is the
$z{-}y$ plane.
The transverse energy densities at a given transverse
position indicated by the dashed line in Fig.~\ref{fig:collision2}
are then given by
\begin{eqnarray}
    \mu_1^3
    &=&
    M_N A_1 \left( \tfrac{4\pi}{3} R_1^3\right)^{-1} \,
    2\gamma_1 \rho_1 \,,
\nonumber \\
    \mu_2^3
    &=&
    M_N A_2 \left( \tfrac{4\pi}{3} R_2^3 \right)^{-1} \,
    2\gamma_2 \rho_2 \,,
\end{eqnarray}
where $A_{1,2}$ are the two nucleon numbers,  $R_{1,2}$ the two nuclear radii,
$\gamma_{1,2}$ the Lorentz factors of each nucleus,
and $y_{1,2}\leq R_{1,2}$ the transverse distances from
the centers of the nuclei in reaction plane
and $x_{1}=x_2\leq R_{1}$ orthogonal to it. We defined $\rho_i \equiv \sqrt{R_i^2-y_i^2-x_1^2}$.
Finally, $y_b \equiv y_1+y_2$ is the impact parameter.

The resulting geometric mean of the scale parameter is
\begin{eqnarray}
    \mu =& \sqrt{\mu_1\mu_2}
    = \Big[ \frac{9M_N^2 A_1A_2}{4\pi^2 R_1^2R_2^2}~ \gamma_1\gamma_2 ~\Big]^{{1}/{6}}\Big[\frac{\rho_1 \rho_2}{R_1 R_2}\Big]^{1/6}
\nonumber \\
   \propto &  \left[ \left(1-\frac{y_1^2+x_1^2}{R_1^2}\right)
     ~\left(1-\frac{(y_b-y_1)^2+x_1^2}{R_2^2}\right) \right]^{{1}/{12}} \,.
\label{eq:geom}
\end{eqnarray}
As shown in Fig.~\ref{fig:Theta},
this function goes to zero as $x_1^2{+}y_1^2\rightarrow R_1^2$,
or $x_1^2{+}y_2^2\rightarrow R_2^2$, so abruptly
that the impact on $v_2$ is negligible. 
\begin{figure}
\includegraphics[width=8.6cm]{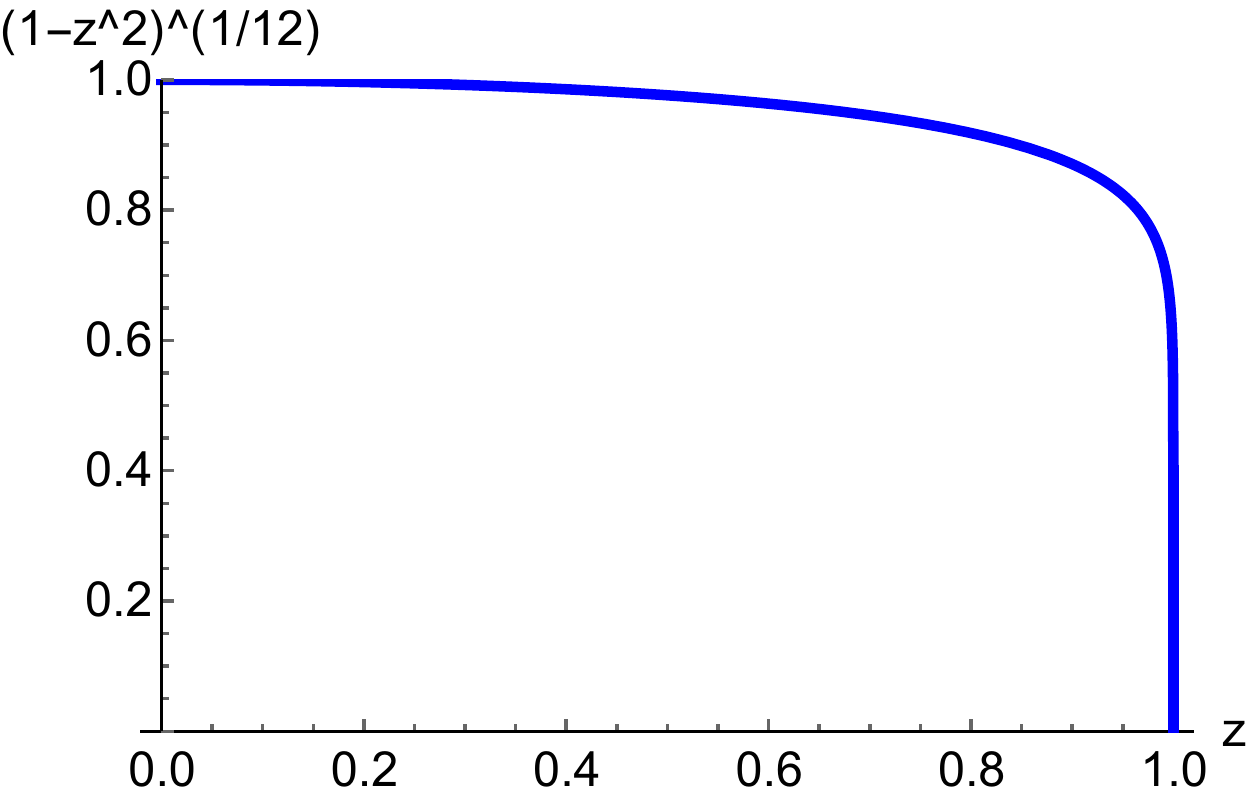}
\caption
    {%
    The function $(1-z^2)^{\frac{1}{12}}$ which,
    in the model (\ref{eq:geom}),
    controls the dependence of the scale $\mu$ on the
    distance from the nuclear boundaries
    }
\label{fig:Theta}
\end{figure}

Let us add that it follows also from the geometric mean of the
energy scales that the difference between
smooth $A+A$ and smooth $p+A$ collisions,
i.e., neglecting effects due to fluctuations, is not dramatic.
For $p+A$ the momentum scale is smaller by a factor $A^{1/6}$
compared to $A+A$ and thus the hydrodynamization time is larger
by that factor 2.5 (for Pb) increasing $0.1-0.2$ fm/$c$ to $0.25-0.5$ fm/$c$.
Thus, also in this case, fluctuations have the strongest
impact on the hydrodynamization time,
implying that $A+A$ and $p+A$ collisions should have similar
properties because, in a holographic description,
the essential difference of incident nucleons versus nuclei
is  solely encoded in their respective energy densities.
We mention in passing that in Ref.~\cite{Waeber:2019nqd} it was 
found that in forward direction of the heavier nucleus in a asymmetric
collision the matter density is larger and thus hydrodynamization happens
slightly earlier.

\section{Entropy production in classical and quantum gauge theories}

It is possible to extend the motivation for the present
contribution to a grander scale. Decoherence, entropy production and
hydrodynamization or thermalization are intensely discussed also in other
fields like quantum gravity and quantum computing. AdS/CFT duality
has the potential to connect all of these fields. It was established
in recent years that there exists an intimate connection to quantum
error correction schemes while by construction AdS/CFT combines quantum
gravity and quantum field theory.

In principle, the connection to QCD and HICs opens the very attractive
possibility for experimental tests of theoretical predictions because the
number of final state hadrons per rapidity interval $dN/dy$ is taken as 
a measure of thermodynamic entropy in the final state after all interactions
have stopped. Obviously this is, however, only helpful if the differences
between CFT and QCD are minor.

As entropy production is equivalent to information loss, this discussion
centers on the question in which sense information can get lost under
unitary time evolution and how this potential information loss in the
boundary theory is related to the generation of the Bekenstein-Hawking
entropy of the formed large black branes in the (AdS$_5\times$S$^5$) bulk.

There exists a fundamental difference with respect to ergodic
properties in the relation between (AdS$_5\times$S$^5$)/CFT and
QCD for high and low temperature, which is most obvious from the
fact that (AdS$_5\times$S$^5$)/CFT is conformal and QCD below the
confinement-deconfinement transition temperature is not. However,
there are strong indications that far above this pseudocritical
temperature the classical solutions of Einstein's equation on
the AdS side produce results that match perfectly relativistic
hydrodynamics which, in turn, provides a near-perfect description of
the experimentally observed properties of the high energy phase of
HICs, see e.g. \cite{vanderSchee:2013pia}. Also, it was shown that on
the string theory side the leading quantum corrections are not large
\cite{Buchel:2008ac,Buchel:2008sh,Stricker:2013lma,Waeber:2018bea}. 
The latter appears to be a general trend. In principle, classical chaos and
quantum chaos can differ fundamentally but it seems that for the high
temperature phase of QCD they do not.

Here we want to address only one aspect of this extensive
discussion to which our calculations may add some insight. For
classical ergodic theories the coarse-grained entropy grows
linearly in time with a rate given by the Kolmogorov-Sinai entropy
$h_{KS}=\sum_i \lambda_i\Theta(\lambda_i)$ defined as the sum of 
all positive Lyapunov exponents $\lambda_i$:
\begin{equation}
dS_{\rm class}=h_\mathrm{KS}~dt 
\label{eq:1}
\end{equation}
Because the definition of entropy is ambiguous in a
non-equilibrium setting one can ask the  question: ``For which definition
of entropy in the quantum theory does one observe linear growth with
boundary time?''. We will not address the much deeper question whether
the condition of linear growth is required or at least well motivated.

Quantum chaos can be described quantitatively in terms of exponential
growth of out-of-time-order correlators (OTOCs)\footnote{
As pointed out by the authors of \cite{Rozenbaum_2019}, the identification of the exponential growth of OTOCs with Lyapunov exponents depends on the specific choice of initial states.
 The general relation between the growth rate of OTOCs and classical Lyapunov exponents  is both nontrivial and so far not fully understood. }
\begin{equation}
  C(t) \sim \Big\langle \left[ \hat W(t),\hat V(0) \right]^2 \Big\rangle ~\sim~\exp(2\lambda_Lt)
\label{eq:3}  
\end{equation}
of suitable operators $\hat W$, $\hat V$ \cite{Larkin,Almheiri:2013hfa,Maldacena:2015waa},
where in the semi-classical limit $\lambda_L$ should be close to the largest classical Lyapunov exponent.  
For many models this behavior was indeed established. For example in Ref.~\cite{Stanford:2015owe} it was found for a weakly coupled matrix $\Phi^4$ theory that 
\begin{equation}
\lambda_L\approx 0.031~\lambda^{3/2}~T
\label{eq:4}
\end{equation}
with the ~$'$t Hooft parameter $\lambda = Ng_{YM}^2$.
Kitaev \cite{kitaev} found for a setting similar to the Sachdev-Ye model \cite{sachdev} that the Maldacena-Shenker-Stanford (MSS) bound
\cite{Maldacena:2015waa} gets saturated for infinite coupling strength:  
\begin{equation}
\lambda_L \to  2\pi T \qquad {\rm for}\quad \lambda \to \infty .
\label{eq:5}
\end{equation}
In \cite{Buividovich:2018scl} the BFSS matrix model was investigated
numerically and it was found, as in all other investigations known to us,
that the leading exponent for a quantum theory stays below the MSS bound,
often even substantially so.

It is indisputable that classical Yang-Mills theories show classical
chaotic behavior, i.e. after an initial phase, which depends on the chosen 
initial conditions, a period of linear growth of the  coarse-grained entropy sets in,
followed by saturation at the thermal equilibrium value. Numerical studies
of classical Yang-Mills theories showing this behavior can be found, e.g.,
in Ref.~\cite{Muller:1992iw,Biro:1994bi,Bolte:1999th,Iida:2013qwa}. 
In \cite{Biro:1994sh} it was conjectured that the largest Lyapunov exponent
in a lattice discretized classical SU(N) Yang-Mills theory at weak
coupling is given by
\begin{equation}
\lambda_L~\approx~0.175 N g_{YM}^2 T = 0.175 \lambda T,
\end{equation}
based on numerical simulations for $N = 2,3$ \cite{Muller:1992iw,Gong:1992yu}. 
In \cite{Kunihiro:2010tg} phenomenological arguments were given that for real QCD 
at the ~$'$t Hooft coupling $\lambda~\approx~11.3$ relevant to the early stage of a HIC, 
the largest Lyapunov exponent is of the order 
\begin{equation}
\lambda_L~\approx~0.3\, T ~\ll~ 2\pi T
\label{eq:6}
\end{equation}  
significantly below the MSS bound.  

There exist many more publications worth mentioning in this context. 
A very recent example dealing with QCD is, e.g.,
\cite{Akutagawa:2019awh}, while in Ref.~\cite{Bianchi:2017kgb} it is
argued that, at least for entanglement entropy, bosons, and a quadratic
Hamiltonian, a linear growth of entropy with the Kolmogorov-Sina\"i
entropy can be derived also for quantum systems.
 Using the coarse graining approach of Husimi distributions, 
the authors of Ref.~\cite{Kunihiro:2008gv} argued that 
the growth rate of the coarse-grained Wehrl entropy of a quantum system
is equal to the Kolmogorov-Sina\"i entropy of its classical counterpart.

Numerical AdS$_5\times$S$^5$ calculations confirm this picture and
add a specific angle. Here, the classical Einstein equation is solved on 
AdS space-time for boundary conditions that mimic, e.g., HICs
\cite{Chesler:2008hg,Chesler:2010bi,Chesler:2013lia,Chesler:2015wra}. 
In this setting the size of the apparent horizon is the appropriate measure
for the produced entropy (see e.g. \cite{Engelhardt:2017aux}). 
We will elaborate on this point in more detail below.\footnote{The 
event horizon is not suitable  as it depends on the entire future history.
} 
Fig. 3 in Ref.~\cite{Chesler:2008hg} shows
for a specific setting that the apparent horizon grows linearly with
time for an appreciable period, just as expected.

Shock wave collisions in $\mathcal{N}=4$ SYM theory, which can be computed 
via their gravitational dual description in AdS$_5$, are believed to behave in many
qualitative aspects very similar to HICs.  
The AdS/CFT duality links the volume element of the apparent  horizon to the entropy density\footnote{Strictly speaking this relation is only valid in the equilibrium case. 
As in e.g. \cite{Wilke_da, Grozdanov:2016zjj} we exploit this relation also off-equilibrium 
and use it to define the entropy(-density) in this case. In general the dictionary giving an interpretation of geometric objects in Anti-de Sitter space in terms of the boundary field theory is far from being complete and remains elusive for many cases, see e.g. \cite{Engelhardt:2017wgc}} 
$s$ times an infinitesimal spatial boundary volume $ d^3x$. 
Using this relation, the longitudinally integrated entropy density in symmetric collisions 
in five dimensional AdS space was calculated, e.g., in Ref.~\cite{Grozdanov:2016zjj} 
(see dashed curves of Fig. 5 in ~\cite{Grozdanov:2016zjj}). 
Our results for the entropy production (longitudinally integrated, given in units of $\mu^2$) 
during symmetric collisions are discussed in Fig. \ref{fig:app_horizon_sym}. 
They are seen to correspond closely to the findings in \cite{Grozdanov:2016zjj}. 
In Fig.~\ref{fig:app_horizon_asym} and Fig.~\ref{fig:comparison} we compare this 
to the analogous computation during asymmetric collisions. 
The code used in this calculation is described in detail in Ref.~\cite{Waeber:2019nqd}.

For large enough times the growth of the area
of the apparent horizon is close to linear, as expected, with a slight
superimposed oscillation which probably averages out over sufficiently
long time periods.\footnote{%
The small wiggling around linear growth seen in
Figs.~\ref{fig:app_horizon_sym} and \ref{fig:app_horizon_asym} can be explained by damped oscillations induced by the lowest quasinormal mode and can thus be expected to fade off for larger times.
} The observation that the rate of growth of
the apparent horizon area is almost identical for the two cases shown
in Figs.~\ref{fig:app_horizon_sym} and \ref{fig:app_horizon_asym} is
relevant for the physics of HICs because the longitudinal thickness of
both ions in the overlap region is very asymmetric in some regions of
the transverse plane. However, the gradient of the linear growth should
be independent of these initial conditions.
\begin{figure}[h]
\includegraphics[width=8.6cm]{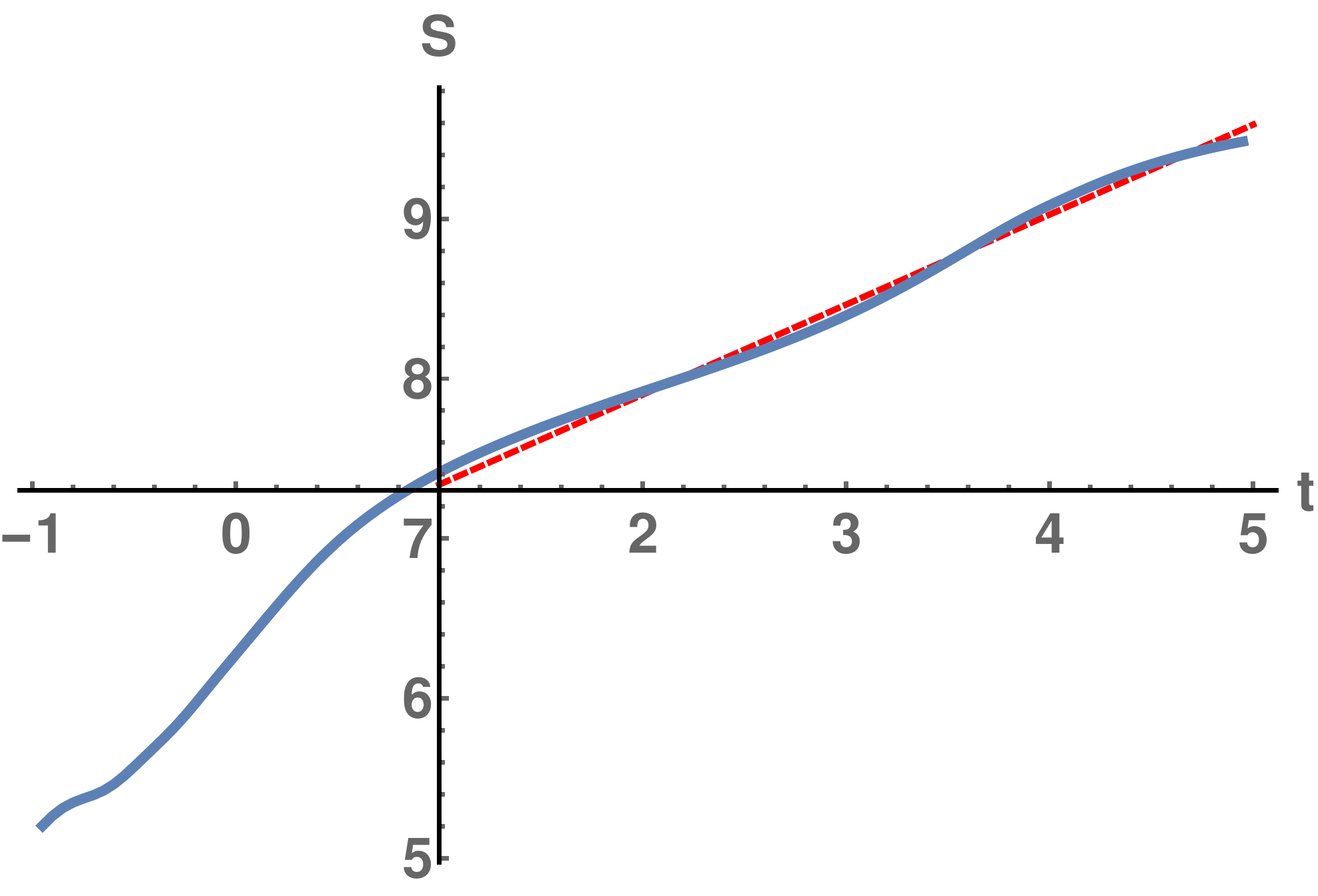}
\caption
{%
 The entropy $S$ (per unit  transverse area on the boundary), produced during a symmetric collision of thin 
 gravitational shockwaves in AdS$_5$ (both shocks have width $ w=0.075/\mu$, 
 where $w$ is the width of the single Gaussian shock waves before the collision), 
 as a function of time $t$, which is given in units of $[t]=[\mu^{-1}]$, where $\mu^3$ 
 is the transverse energy density of the shock fronts. The gauge/gravity duality relates 
 the entropy density $s$ to the volume element of the apparent horizon. 
 To estimate the entropy production we integrate over the longitudinal coordinate. 
 $S$ is given in units of $\mu^2$.  For large enough times linear growths seems to be 
 a good approximation. The shock fronts touch at $\mu t=0$. The linear fit, plotted as 
 a red dashed line, is included to guide the eye. Due to the finitely sized spatial box, 
 in which we study the gravitational collision, we could not follow the time evolution 
 long enough to observe a potential saturation regime for the entropy (see e.g. \cite{Tsai:2010yb}).} 
\label{fig:app_horizon_sym}
\end{figure}

\begin{figure}[ht]
\includegraphics[width=8.6cm]{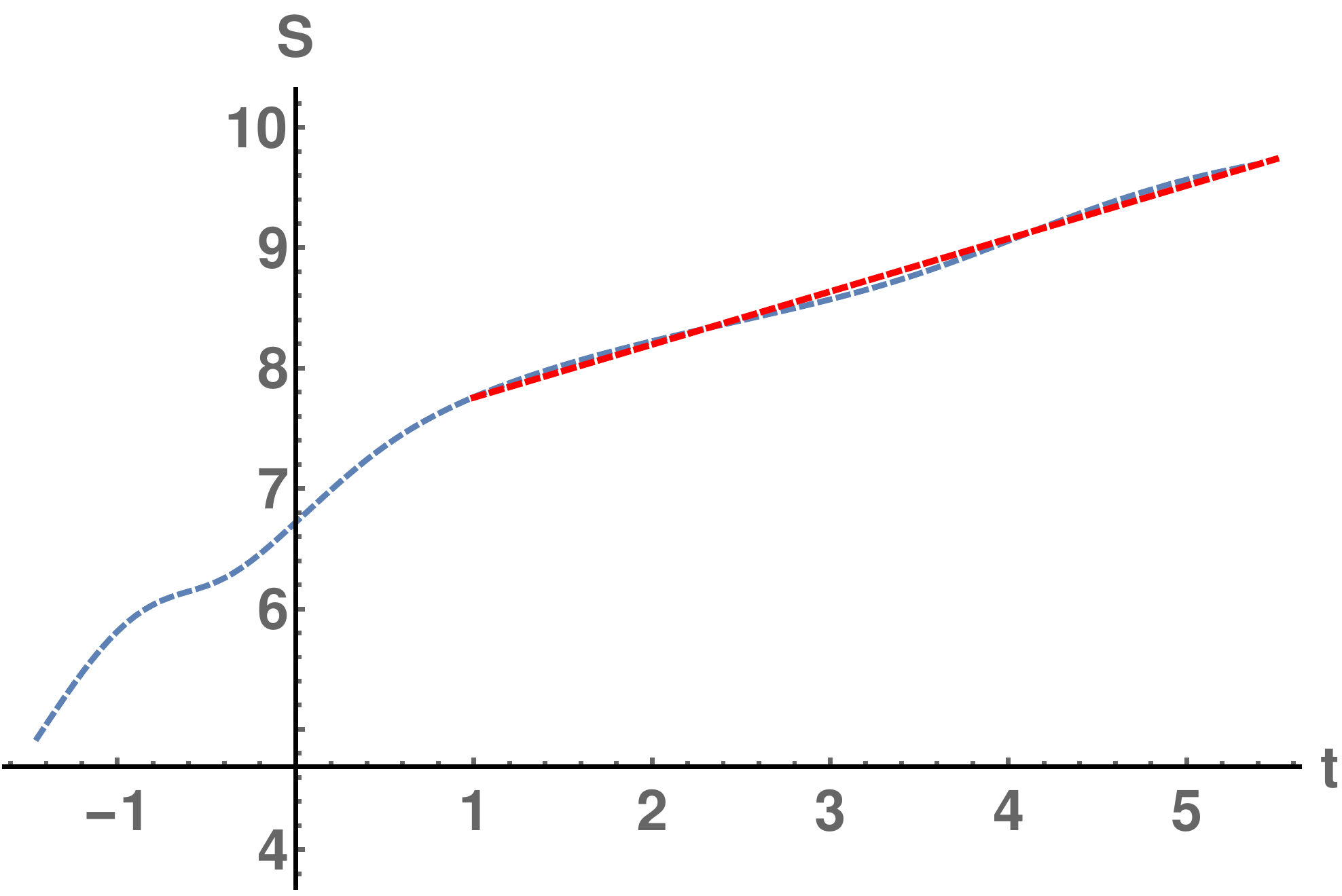}
\caption
{%
  The analogous plot as shown in Fig.\ref{fig:app_horizon_sym} but for an asymmetric collision of shock waves with widths $w_+=0.075/\mu$ (right moving) and $w_- = 0.25/\mu$ (left moving).}  
\label{fig:app_horizon_asym}
\end{figure}

\begin{figure}[ht]
\includegraphics[width=8.6cm]{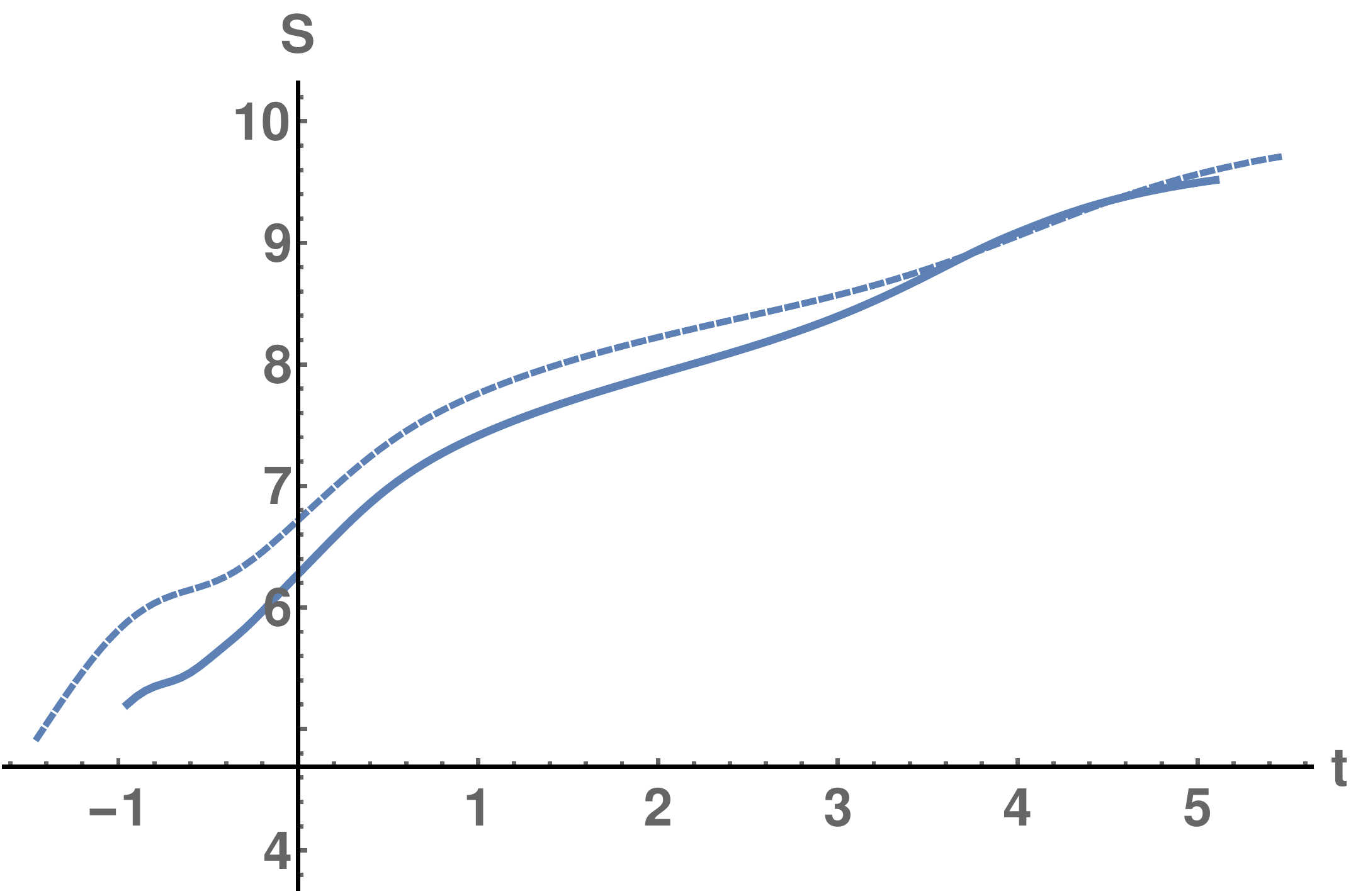}
\caption
{%
  Comparison of the symmetric (red, dashed, Fig.\ref{fig:app_horizon_sym}) 
  and asymmetric (gray blue, Fig.\ref{fig:app_horizon_asym}) cases.}  
\label{fig:comparison}
\end{figure}

Thus, at large times linear entropy growth does not only seem to connect
classical and quantum chaos for QFTs but also their holographic dual
description. Obviously, these similarities could well be accidental
at our present stage of understanding, but they are sufficiently intriguing}  
to warrant further research.

\section{Conclusion}
In this note we have argued that detailed holographic calculations
for asymmetric collisions are highly relevant for any quantitative
description of realistic HICs. We have performed such calculations and
have found that:
\begin{itemize}
\item
The characteristic large fluctuations in transverse energy and entropy
densities, which are required in hydrodynamic descriptions to explain
the observed large event-by-event fluctuations of flow observables,
delay hydrodynamization and equilibration in the holographic description so strongly that they
are still significant at time of 1--2 fm/$c$ when hydrodynamics becomes
definitively applicable.
\item
In contrast, the effect of the transverse dependence of energy densities
in peripheral collisions has only a minor impact on the hydrodynamization
time, such that it is well motivated to initialize hydrodynamics for the
entire system simultaneously.
\item
p+A and A+A collisions do not show major differences in hydrodynamization properties.
\item
The long time linear growth of the apparent horizon is very similar for
symmetric and asymmetric collisions which supports its interpretation
as entropy.
\end{itemize}  
 Finally, it should be noted that while the delayed hydrodynamization and equilibration  caused
by initial state fluctuations is helpful to explain the observed $v_3$ fluctuations 
it also increases concerns that thermalization might not happen as rapidly 
as is usually assumed in the interpretation of the medium modifications of 
``hard'' probes.

\section*{Acknowledgements}

 BM acknowledges support from the U.S. Department of Energy 
grant DE-FG02-05ER41367.
LY acknowledges support from the U.S. Department of Energy 
grant DE-SC\-0011637.  SW acknowledges support from the Elite Network of Bavaria in form of a  Research Scholarship grant. AS and SW acknowledge support by BMBF (Verbundprojekt 05P2018,
05P18WRCA1).

\end{document}